\documentstyle[12pt]{article}

\newcommand{\be}{\begin{equation}}
\newcommand{\ee}{\end{equation}}
\newcommand{\bea}{\begin{eqnarray}}
\newcommand{\eea}{\end{eqnarray}}
\newcommand{\ci}{\cite}
\newcommand{\bi}{\bibitem}

\newcommand{\dd}{\partial}

\newcommand{\half}{\frac{1}{2}}

\def\dal{\,\lower0.3ex\vbox{\hrule\hbox{\vrule\kern2pt\vbox{\kern4pt\kern4pt}
\kern2pt\vrule}\hrule}\,}

\begin{document}

\title{{\bf SOLITON TUNNELING}}
\vspace{1 true cm}
\author{G. K\"ALBERMANN \\
Faculty of Agriculture \\
and \\
Racah Institute of Physics \\
Hebrew University, 91904 Jerusalem, Israel \\}

\maketitle

\begin{abstract}
\baselineskip 1.5 pc
We present a numerical simulation of the scattering of a topological soliton 
off finite size attractive impurities, repulsive impurities 
and a combination of both.
The attractive and attractive-repulsive cases show similar features
to those found for $\delta$ function type of impurities.
For the repulsive case, corresponding to a finite width barrier,
the soliton behaves completely classically. No tunneling occurs for sub-barrier
kinetic energies despite the extended nature of the soliton.
\end{abstract}

\newpage
\baselineskip 1.5 pc

Topological solitons are frequently mentioned as possible candidates for the 
description of particles. Very notably the skyrmion \ci{sk1,an} has been
proposed as a sound model of the nucleon. The topology of the skyrmion serves
as a classical picture of the baryon current. 
It is quite clear that if nucleons can be considered as solitons in a nonlinear
chiral lagrangian, 
their behavior in nuclei has to follow from the same framework. In particular
one of the most intriguing characteristics of the quantum mechanical
behavior of nucleons in nuclei is the tunneling through a barrier.
The quantum picture of the nucleon as a wave allows for clear predictions of
the tunneling rates for sub-barrier energies. 
The question then arises as to what would be the behavior of solitons
colliding with a barrier in similar circumstances. In particular a soliton
model can provide some partial answers on the longstanding problems of the
tunneling times around which there is much controversy in the literature
\ci{rec}.
If the soliton is to behave as a classical particle, then there can not be
sub-barrier tunneling at all. However, for an extended object the answer is not
so straightforward ( recall a high jumper whose center of mass goes
through the barrier, while the jumper glides above it).
\baselineskip 1.5 pc
The simplest case of such a process would be a one dimensional collision of
a topological soliton -like the kink or the sine-gordon soliton- with a barrier. 
Such processes can be catalogued under the title of soliton-impurity 
interactions. 

Some time ago Kivshar et al.\ci{kiv} investigated the 
scattering of a kink and a sine-Gordon soliton off an attractive $\delta$
function well and found extremely interesting results, such as the existence
of resonant behavior of the soliton, trapping, reflection and excitation of
the so-called impurity modes.
The use of a $\delta$ function impurity allowed them to predict analytically
the existence of
windows of reflection in between trapping and resonances regions as a function
of impinging velocity, defying a classical interpretation
as particle behavior. They did not consider a finite well nor the
barrier case. The results found for the sine-Gordon and the kink models were
essentially the same. Extensions of the model to include inhomogeneities
were undertaken in ref.\ci{cuba}.
An investigation of the chaotic behavior of the
residence time of the soliton inside an attractive impurity as a function of initial
location was performed by Fukushima and Yamada \ci{fuk}.

In the present work we calculate numerically the interaction of a kink with
a finite width impurity of the attractive, repulsive, or mixed case.
The basic model is decribed by the lagrangian
\bea \label{lag}
 {\cal L} & = &\half \dd_\mu\phi\,\dd^\mu\phi+
{1 \over 4 }\Lambda~{\bigg(\phi^2 - {m^2\over \lambda}\bigg)}^{2}
\eea
Here 
\be \label{Lambda}
\Lambda = \lambda + U(x)
\ee
$\lambda$ being a constant, and $U(x)$ the impurity potential,

\be \label{u}
U(x) = h_1~{cosh(\frac{x-x_1}{a_1})}^{-2} + h_2~{cosh(\frac{x-x_2}{a_2})}^{-2}
\ee
allowing a combination of both repulsive $h_1 > 0$ and attractive $h_2 < 0$
impurities. 

The partial differential equations of motion were solved using a finite
difference method checked against the results of Kivshar et al. \ci{kiv} 
(although we do not agree entirely 
with the actual values of the final velocities quoted there)
and the free analytical solution.
We took a soliton initially at x = -3 shot to the
right with initial velocity $v$ onto an impurity located at x = 3.
The spatial boundaries were taken to be $ -40 < x < 40$ , with a grid of
$dx = 0.04$ and a time lapse of $dt= 0.02$ up to a maximal time of $T= 200$
(10000 time steps). This choice proved efficient in preventing 
numerical instabilities and still not exceedingly time consuming.
The upper time limit allows for resonant passes to decay and permits a
clear definition of the asymptotic behavior of the soliton. Care has to be 
taken not to exceed a certain time limit in order to prevent reflection from
the boundaries. The asymptotic velocities, for the reflected and transmitted 
cases were calculated using the actual motion of the center of the soliton and
with the theoretical expressions for the kinetic and potential energies of the
free soliton.

We chose the parameter $\lambda = m^{2} $ in eq.~(\ref{lag}) without loss of
generality and allowed three different values for $m = 0.7 , 1, 1.5$
so chosen in order to study solitons whose effective widths $\approx 
{\displaystyle{1\over m}}$
are bigger, comparable, and smaller than the barrier widths $\approx 
{\displaystyle{a\over 6}}$ 
, where $a$ is the parameter in the argument of U(x) in eq.~(\ref{u}).
For that purpose we took a repulsive barrier whose width is fixed at
$a_1 = 1$ and an attractive barrier with $a_2 = 0.3$. The reason
for this distinction was biased by our knowledge 
of the nuclear potential, that for
heavy nuclei $\alpha$ decay has a deep and short range attractive
well and much broader repulsive barrier generated by the Coulomb interaction.
The choice of potential heights was determined by the desire to see
all the effects in a range of reasonable velocities ( not too low nor to high)
around $ v \approx 0.25$. Trial and error and the above considerations
lead us to choose $h_1 = 1$ and $h_2 = -6$. 
The lack of analytical solutions for finite size barriers prevented
us from general predictions and we therefore limited ourselves to
the above parameter set.

Figure 1 shows the impinging soliton as well as the various barriers.
Figures 2-5 show the final velocity $v'$ as a function of initial
velocity $v$ for the repulsive $h_1 =1$,  $h_2 =0$ , attractive $h_1 = 0$,
$h_2 = -6$, attractive-repulsive and repulsive-attractive cases respectively.
The repulsive case of Fig.2 shows a clear particulate behavior. 
The soliton is reflected,
$v' < 0$, up to a certain speed for which the effective barrier height 
becomes comparable with the kinetic energy and then there is a sudden jump
to transmission. In all three cases the transmission starts at the same
kinetic energy, with minor differences due to the effective barrier that is 
composed of the kink and the barrier.
The attractive case of Fig. 3 is analogous to the $\delta$ function pattern
found by Kivshar et al. \ci{kiv}. There are islands of reflection
in between trappings and resonant behavior for which the soliton remains
inside the impurity and oscillates exciting the so-called impurity mode.
Again the higher the mass, the smaller the critical velocity for which
transmission starts.
The details of the reflection islands depend strongly on the parameters
but the general trend is analogous for all three mass cases.
Figure 4 depicts the results for a combination of attractive and repulsive 
impurities.
For low velocities reflection dominates -induced by the repulsive
impurity-, then trapping and resonant behavior occurs with islands of reflection
followed by transmission essentially dictated by the same impurity (Compare
to Fig. 2).
The repulsive-attractive case of Fig. 5 is similar to the repulsive case
for velocities below transmission and the critical speed is here determined
mainly by the attractive impurity that can drag back the soliton after is passes
through the barrier. It appears that the larger the mass (the thinner the
soliton) the attractive impurity is more capable of trapping, thereby 
producing a somewhat counterintuitive behavior for which the larger mass solitons
tend to need a higher initial velocity in order to traverse them.

Concerning the permanence time inside the barrier, there is always
a time delay in the impurities, in contradistinction to the quantum-mechanical
Hartmann effect \ci{har}. Also, energy is conserved in the scattering.

The present investigation addressed the one dimensional case. 
In order to relate more closely to actual nuclear ( or optical) tunneling 
phenomena one has to consider higher dimensions, such as the $O(3)$ two-
dimensional case or the skyrmion, including eventually rotations of the soliton
and other effects, like fluctuations. Moreover, actual nuclear barriers are 
dynamical and not stiff. There is then a need to allow for more flexibility
in the impurities as well as the possiblity of dissipation.

\vspace{3 pc}

{\bf Acknowledgements}

It is a pleasure to thank Prof. Ignatovich from Dubna 
for a motivating question that triggered this project.

\newpage
{\bf Figure Captions}

\begin{enumerate}
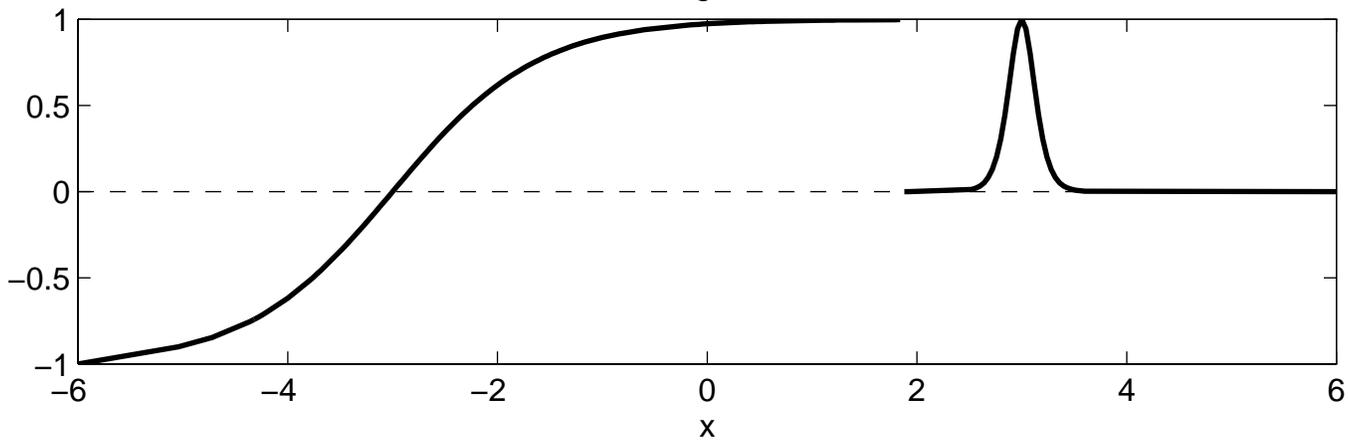
\item[{\bf Fig. 1}:]  From top to bottom: Kink with $ m= 1$ 
impinging from the left
onto a repulsive barrier. kink with $ m= 1$ 
impinging from the left onto an attractive impurity.
Kink with $ m= 0.7$ impinging from the left
onto an attractive-repulsive system.
Kink with $ m= 1.5$ impinging from the left
onto a repulsive-attractive arrangement.

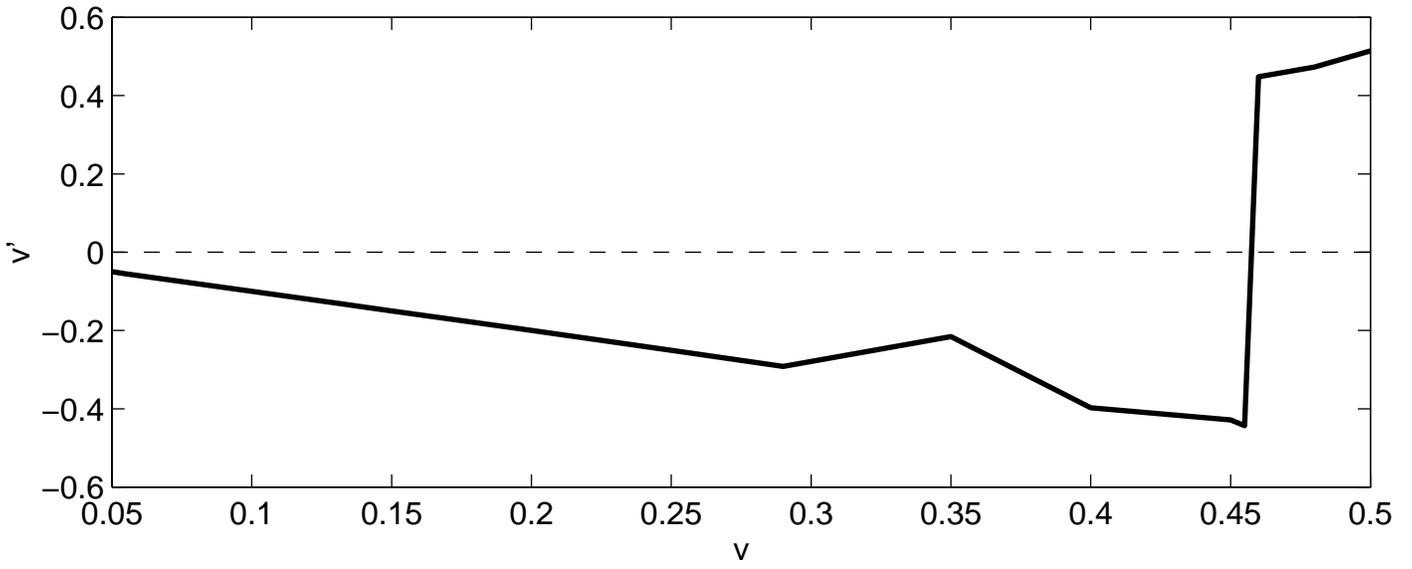
\item[{\bf Fig. 2}:] Final velocity $v'$ as a function of the initial
velocity $v$ for soliton masss parameters $m = 0.7$ upper curve $m = 1$
middle curve and $m = 1.5$ lower curve for the repulsive barrier.

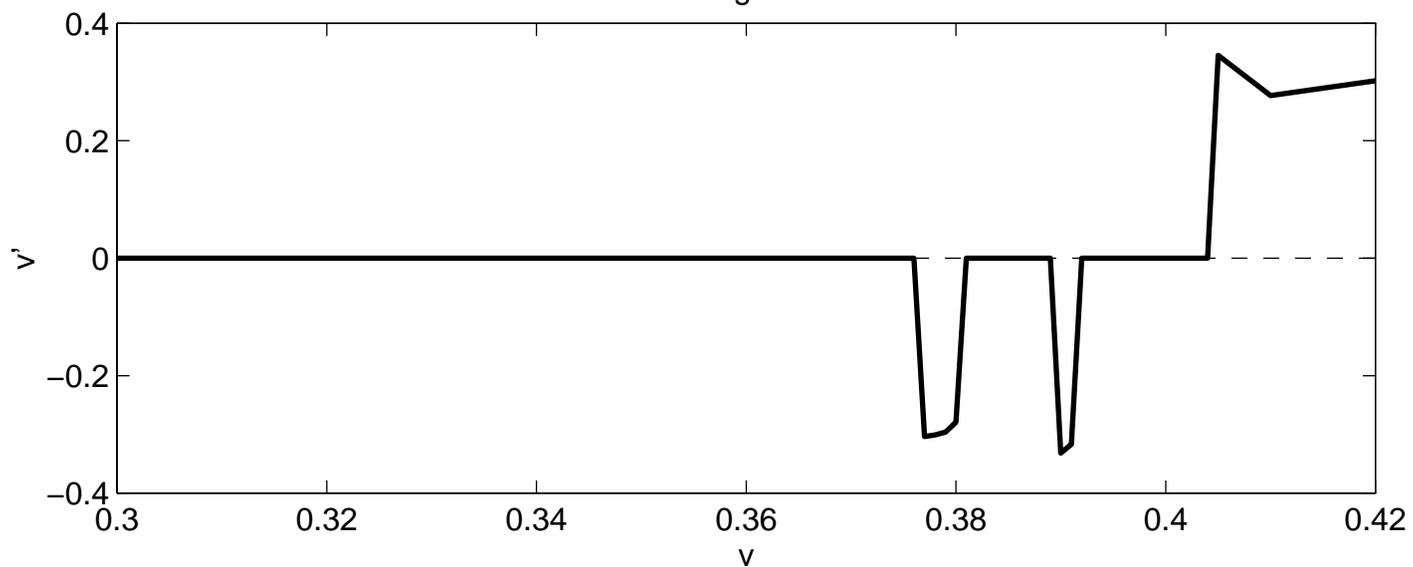
\item[{\bf Fig. 3}:] Same as figure 2 for the attractive case.
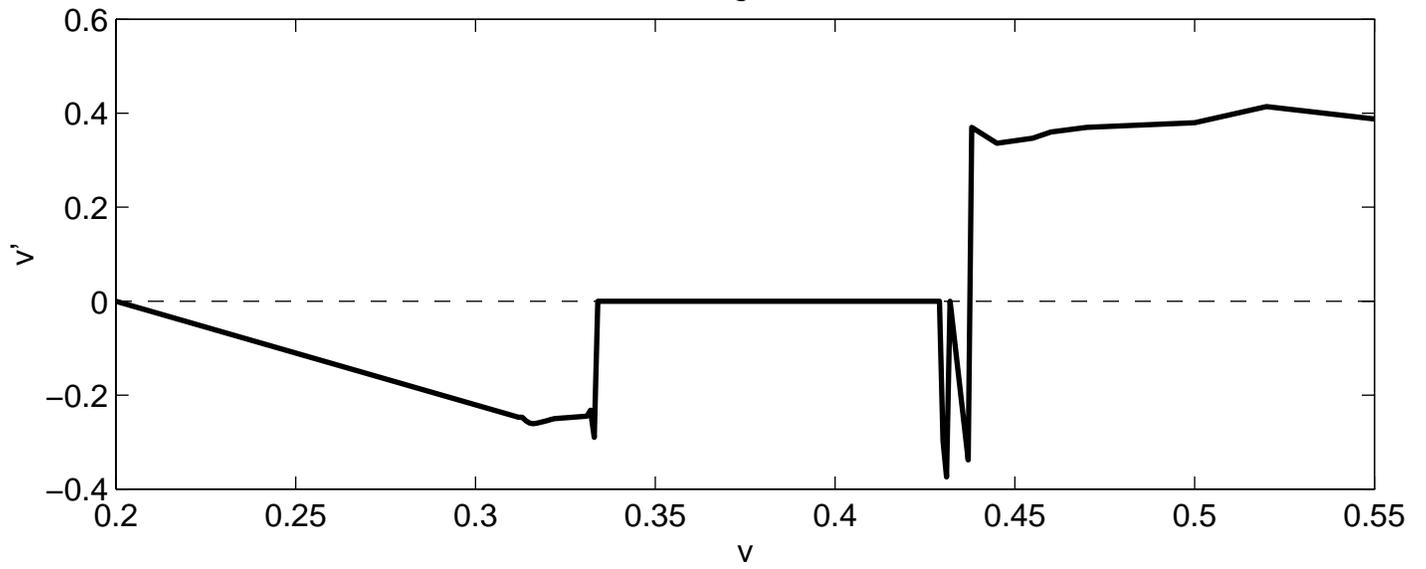
\item[{\bf Fig. 4}:] Same as figure 2 for the attractive-repulsive case.
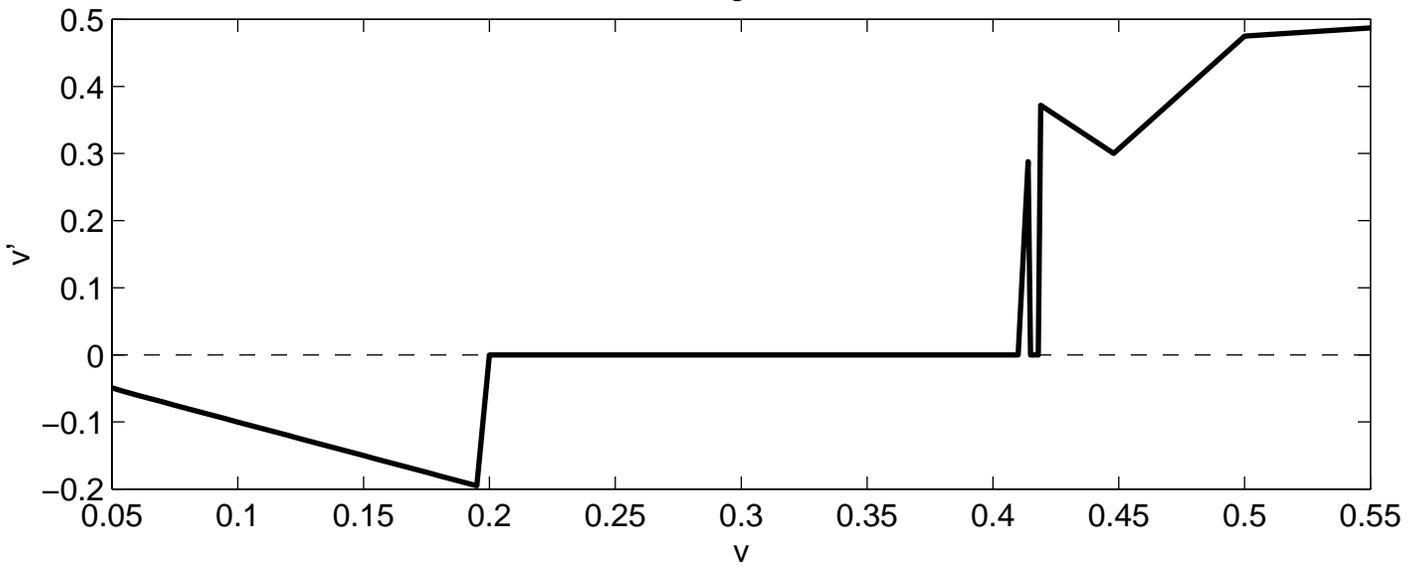
\item[{\bf Fig. 5}:] Same as figure 2 for the repulsive-attractive case.
\end{enumerate}
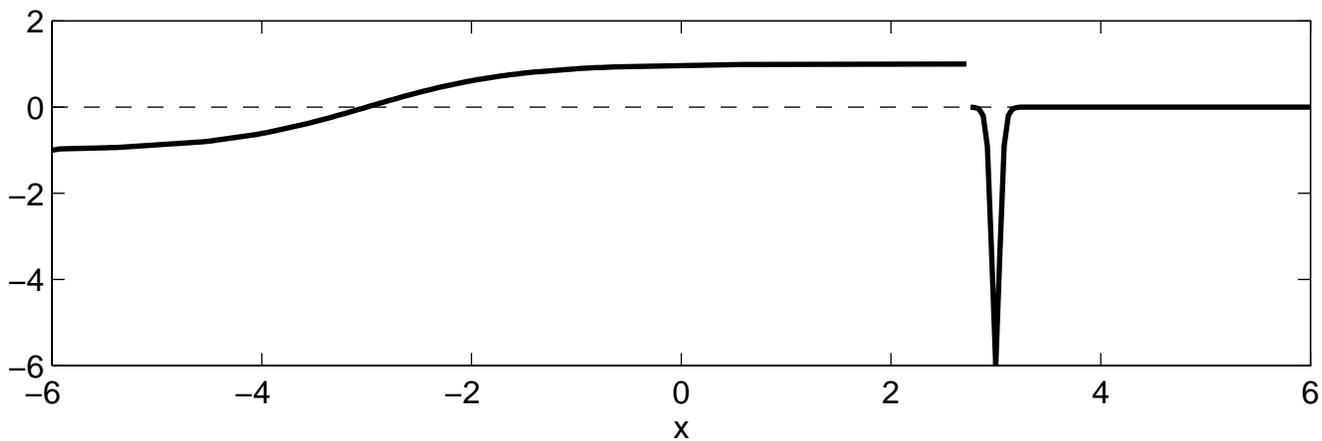
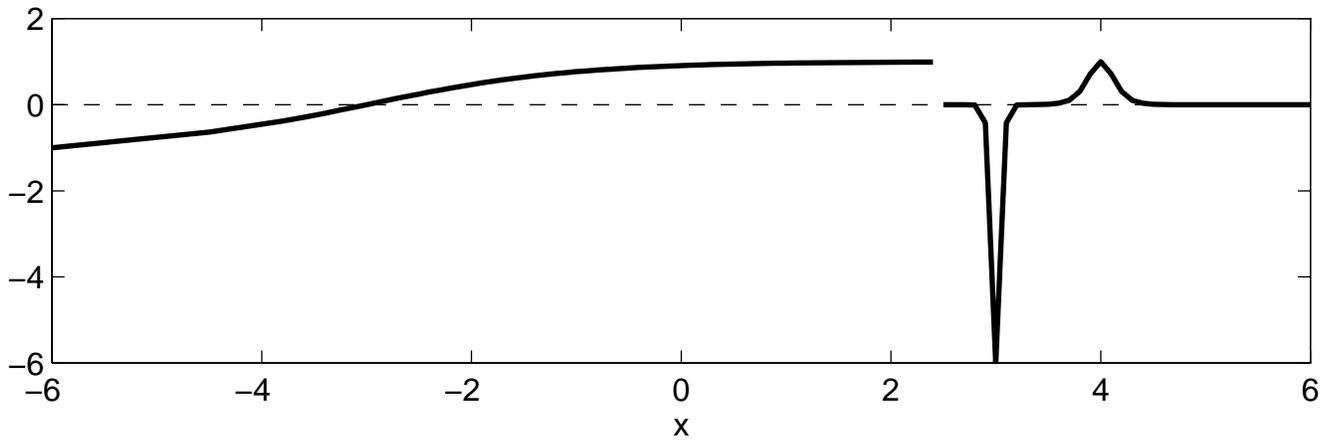
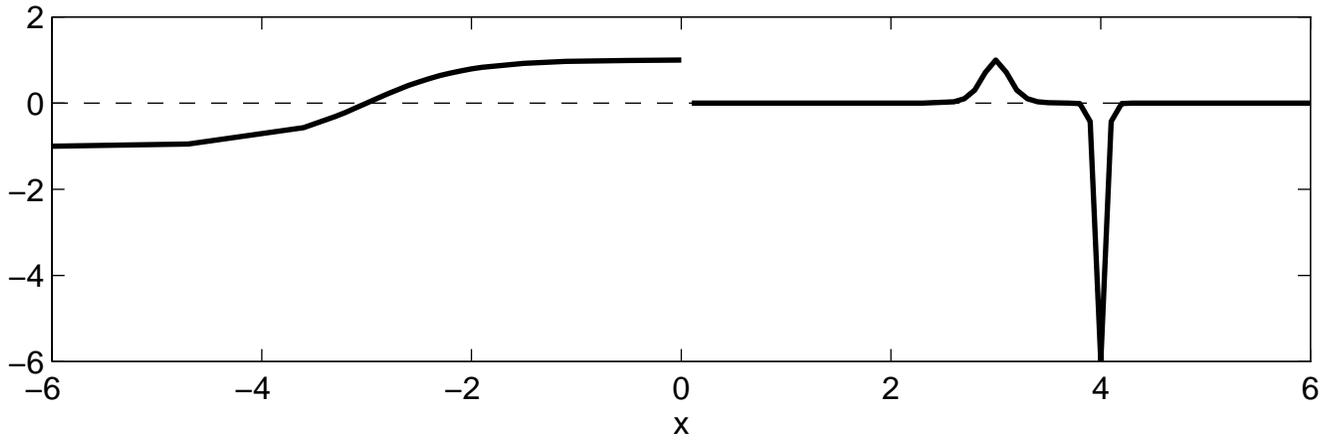
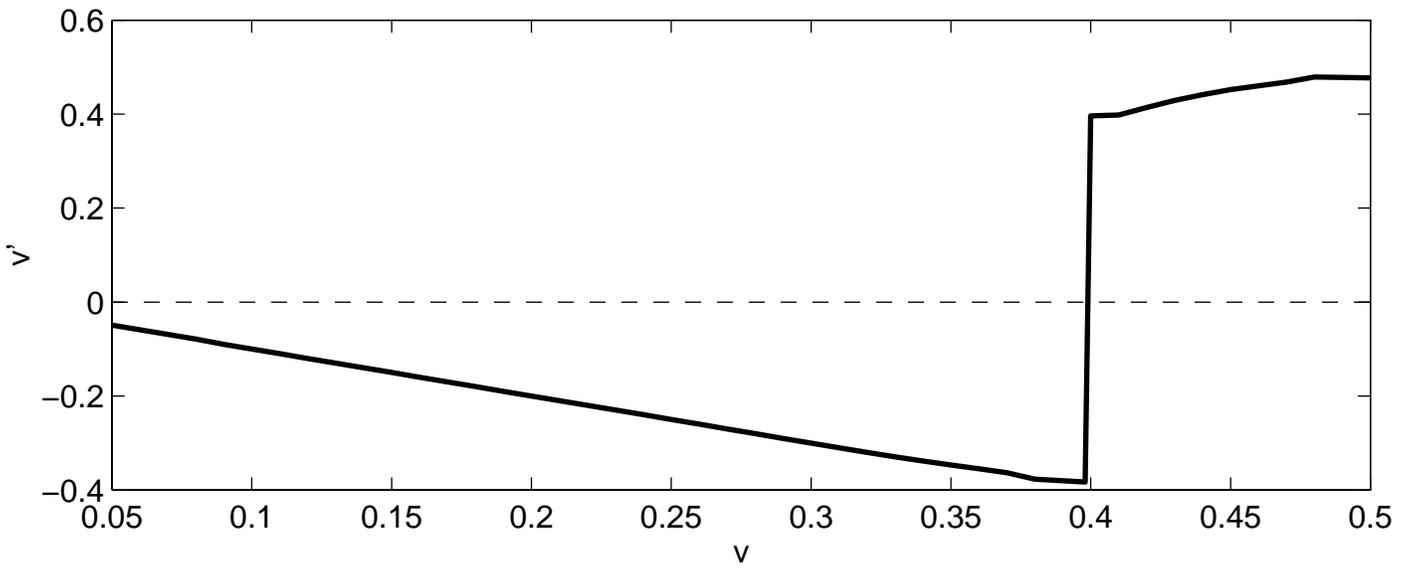
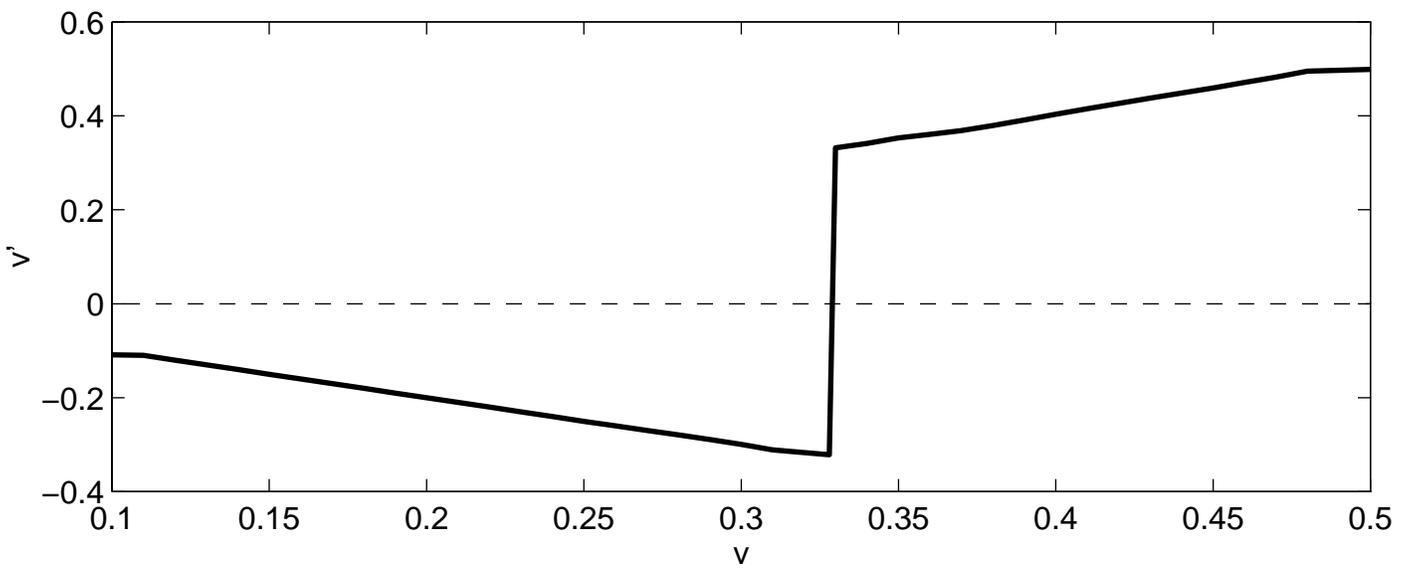
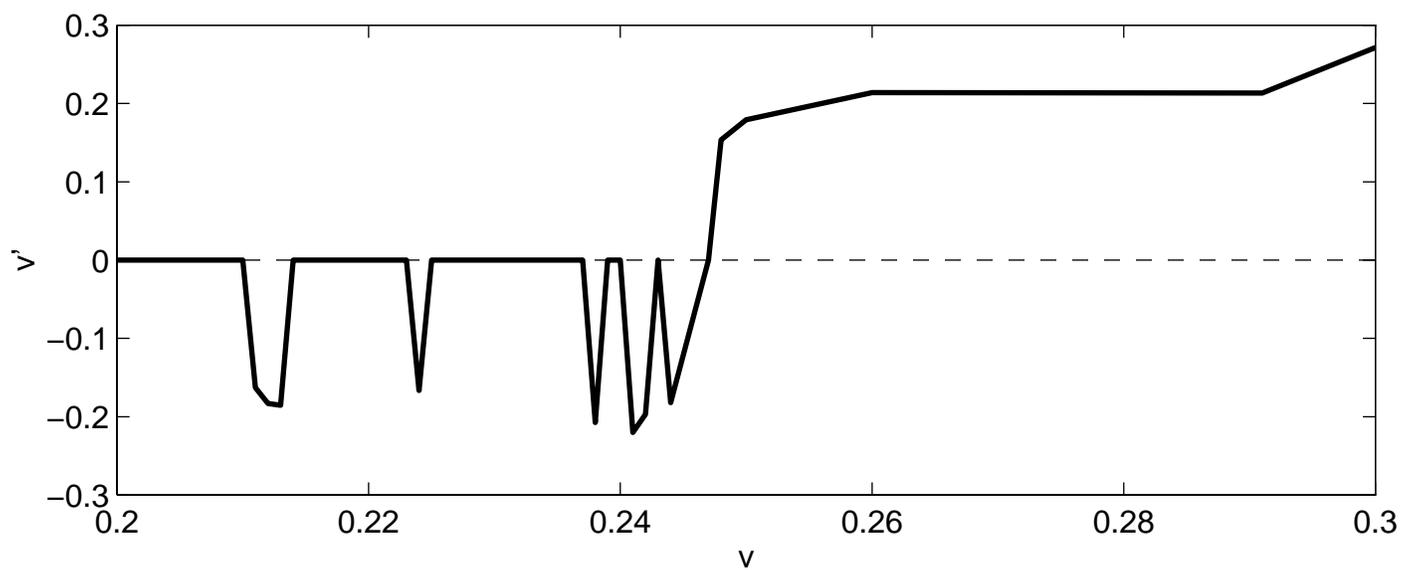
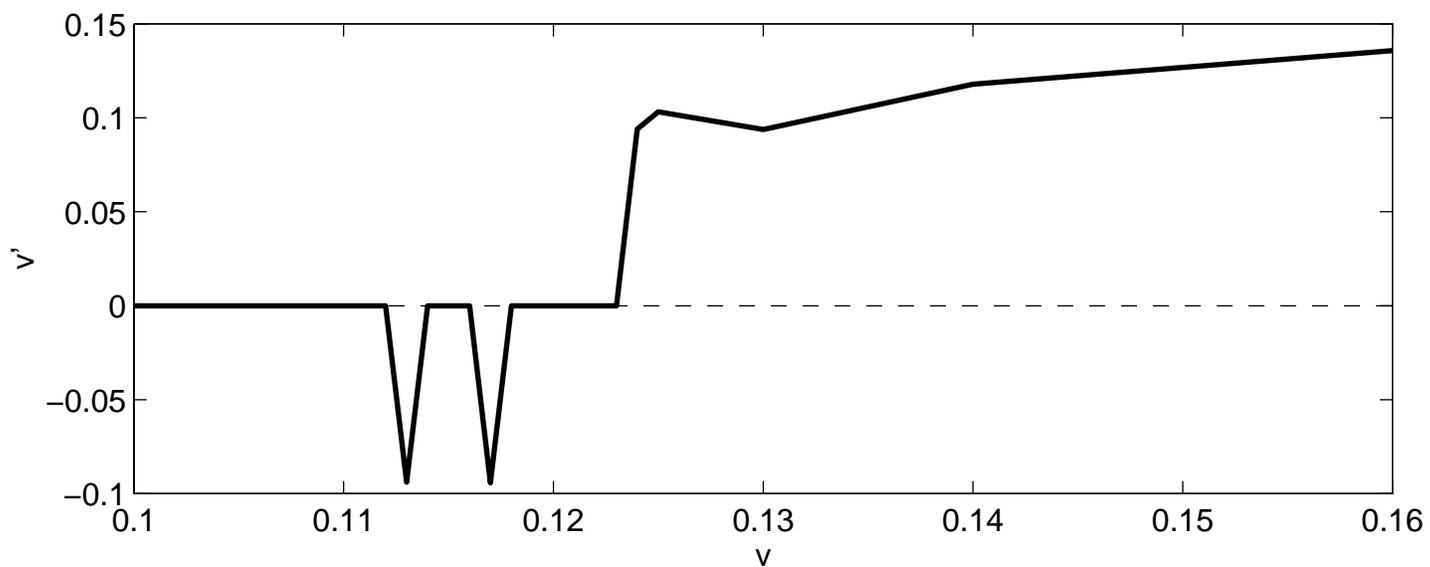
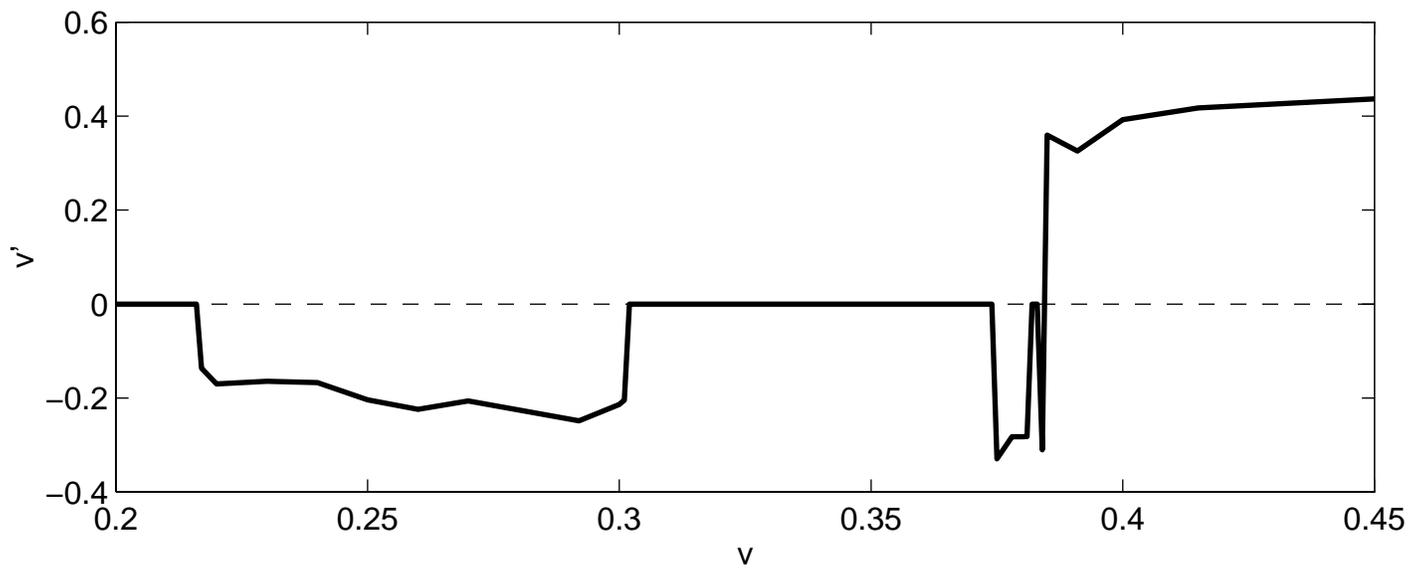
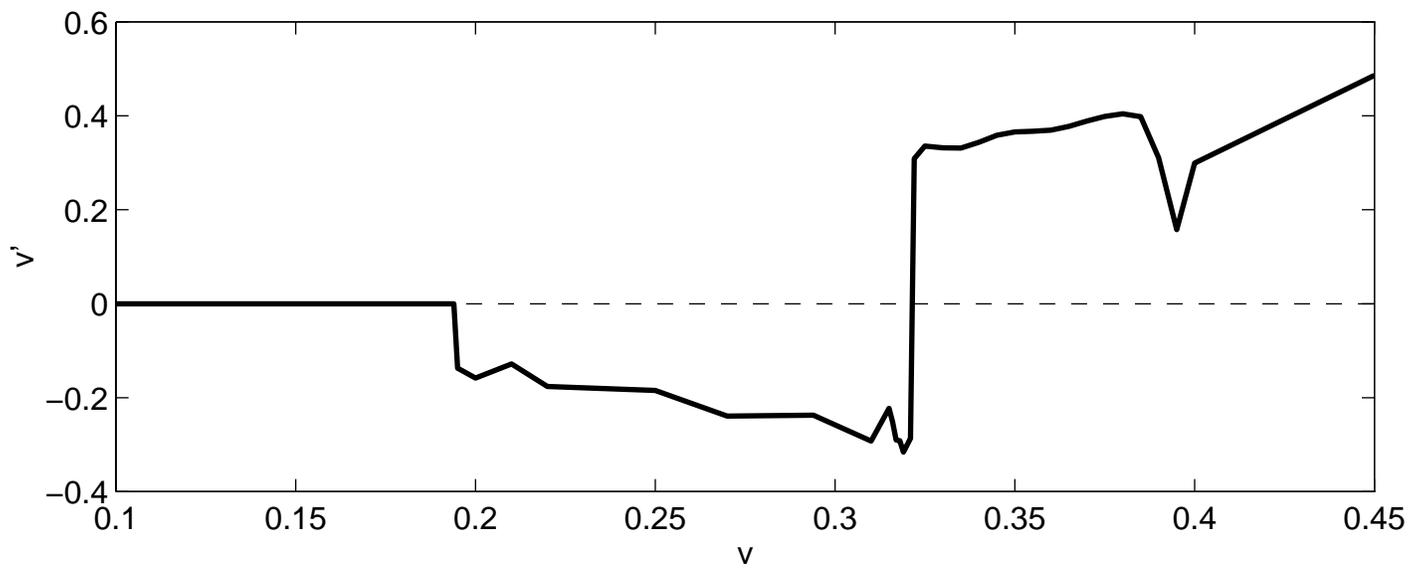
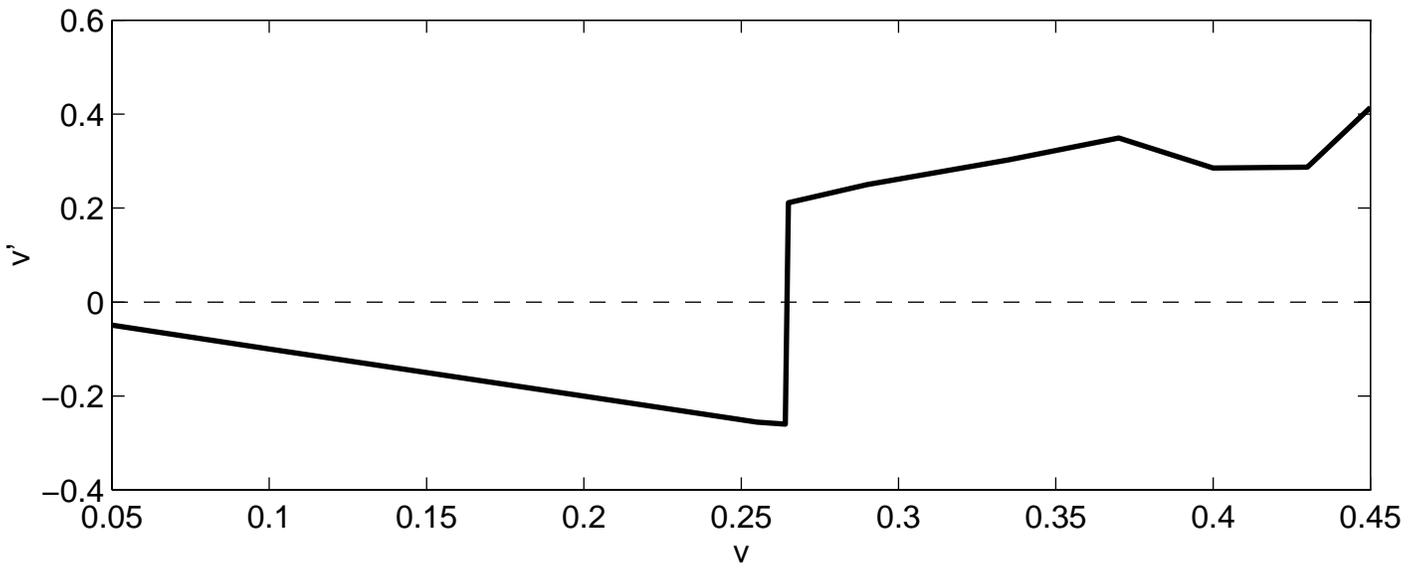
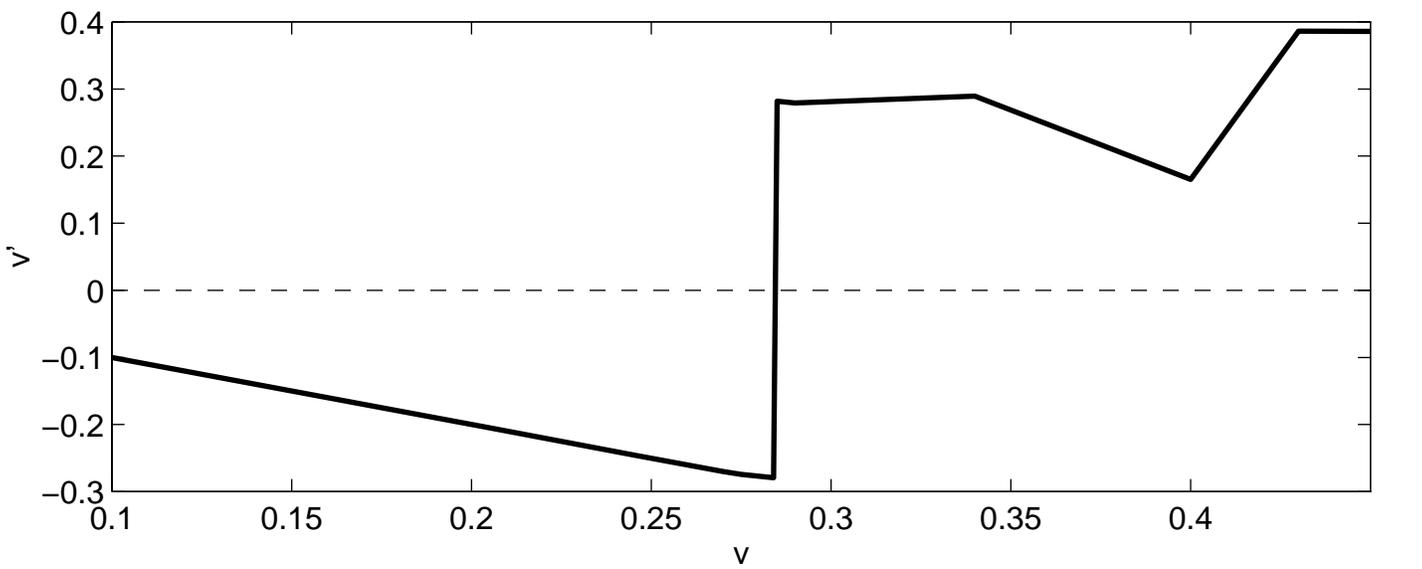
\end{document}